\documentstyle[12pt]{article}
\newcommand{\beq}{\begin{equation}}
\newcommand{\eeq}{\end{equation}}
\newcommand{\beqa}{\begin{eqnarray}}
\newcommand{\eeqa}{\end{eqnarray}}

\begin{document}

\pagestyle{empty}

\hfill TK 95 29

\smallskip

\begin{center}

{\large { \bf THE REACTION $\pi N \to \pi \pi N$ At THRESHOLD}}

\vspace{1.cm}

Ulf--G. Mei{\ss}ner\\
{\it Universit\"at Bonn, ITKP, Nussallee
14-16, D--53115 Bonn, Germany}

\vspace{0.4cm}

\end{center}

\baselineskip 10pt

\noindent I summarize the results of the complete one--loop chiral perturbation
theory calculation performed recently. It is shown that it allows to accurately
pin down the  isospin two, S--wave $\pi \pi$ scattering length $a_0^2$. On the
other hand, interesting resonance physics makes a precise determination of
$a_0^0$ very difficult.

\vspace{0.5cm}

\baselineskip 14pt

\noindent {\bf 1 $\quad$ PRELIMINARIES}

\vspace{0.3cm}

\noindent The interest in the reaction $\pi^a (k) + N(p_1) \to
\pi^b (q_1) + \pi^c (q_2) + N(p_2)$, where '$a,b,c$' are pion isospin indices
and $N$ denotes the nucleon (neutron or proton),
stems mostly from the fact that it includes (besides many other contributions)
the one--pion exchange diagram with the four--pion vertex. This allows to
extract information about the on-shell low--energy $\pi \pi$ interaction
which is the purest testing ground of our understanding of the spontaneous
and explicit chiral symmetry breaking in QCD. The relation between the
threshold $\pi \pi N$ and the threshold $\pi \pi$ amplitudes has until
recently only be known at lowest order (tree  level).
To precisely determine the $\pi \pi$ S--wave scattering lengths
$a_0^I$ (with $I=0$ or $2$), a one--loop calculation within the framework
of heavy nucleon chiral perturbation theory is called for.
I will present the most salient results of a recent investigation
of that type here (for details, see Ref.[1]).

At threshold, the on--shell amplitude in the $\pi^a N$ cms can be expressed in
terms of two threshold amplitudes, called $D_1$ and $D_2$,
\beq
T = i \vec{\sigma} \cdot \vec{k} \, \biggl[ D_1 \, \bigl( \tau^b \delta^{ac}
+ \tau^c \delta^{ab} \bigr) + D_2 \, \tau^a \delta^{bc} \biggr] \, \, ,
\eeq
where $\vec{\sigma} $ denotes the spin vector of the initial nucleon. The
chiral expansion of the amplitudes $D_{1,2}$ takes the form
\beq
D_{1,2} = (f_0)_{1,2} + (f_1)_{1,2} \, \mu + (f_2)_{1,2} \, \mu^2
 + \dots \, , \quad
\mu \equiv M_\pi / m \, \, ,
\eeq
modulo logs and $M_\pi$ ($m$) denotes the pion (nucleon) mass.

\vspace{0.5cm}

\noindent {\bf 2 $\quad$ LOW--ENERGY THEOREMS FOR $\pi N \to \pi \pi N$}

\vspace{0.3cm}

\noindent In Ref.[2], the first two terms
of the chiral expansion, Eq.(2), were
calculated. This amounted to a set of novel low--energy theorems (LETs)
since in the resulting expressions only well--known parameters appear,
\beq
D_1 = {\cal C} \biggl[ 1 + \frac{7}{2} \, \mu \biggr] =
2.4 \, \, {\rm fm}^3 \, , \quad
D_2 = -{\cal C} \biggl[ 3 + \frac{17}{2} \, \mu \biggr] =
-6.8 \, \,  {\rm fm}^3 \, \, ,
\eeq
with ${\cal C} = g_{\pi N}/ (8 m F_\pi^2)$. In Refs.[2,3], the threshold data
(i.e. for pion kinetic energies less than 30 MeV above threshold)
for the two channels $\pi^+ p \to \pi^+ \pi^+ n$ (which is sensitive to $D_1$)
and $\pi^- p \to \pi^0 \pi^0 n$ (which is sensitive to $D_2$) were fitted,
\beq
D_1^{\rm exp} = 2.26  \pm 0.12 \, \, {\rm fm}^3 \, , \quad
D_2^{\rm exp} = -9.05 \pm 0.36 \, \, {\rm fm}^3 \, \, \, ,
\eeq
which shows the expected pattern of deviation from the LETs,
namely small/sizeable for the isospin two/zero $\pi \pi$ final state (since
$D_1$ is directly proportional to the $I_{\pi\pi}=2$ amplitude). Notice that
more global fits including data up to much higher energies tend to give
somewhat different values for $D_{1,2}^{\rm exp}$ because the threshold region
does not have sufficient statistical weight in such global fits (see e.g.
Ref.[4]). At this order, however, the $\pi \pi$ interaction to be extracted
does not come into play, i.e. one has to go one order higher and calculate
{\it all} terms of order $M_\pi^2$.

\vspace{0.5cm}

\noindent {\bf 3 $\quad$ OUTLINE OF THE CALCULATION TO ORDER $M_\pi^2$}

\vspace{0.3cm}

\noindent Here, I can just give a flavor of the rather involved calculation
to ${\cal O}(M_\pi^2)$ presented in Ref.[1]. There are essentially four types
of contributions which have to be considered. These are related to the
effective pion--nucleon Lagrangian ${\cal L}_{\pi N} = {\cal L}_{\pi N}^{(1)}
+ {\cal L}_{\pi N}^{(2)} + {\cal L}_{\pi N}^{(3)}+ {\cal L}_{\pi \pi}^{(2)}
+ {\cal L}_{\pi \pi}^{(4)}$ as follows (where the superscript $'(i)'$ gives the
chiral dimension):

\begin{itemize}

\item[$\bullet$] {\it One--loop graphs with insertions from
${\cal L}_{\pi N}^{(1)}$ and ${\cal L}_{\pi \pi}^{(2)}$ }:
Including mass and coupling constant
renormalization, there are $36 \times 2$ topologically different one loop
graphs to be considered (the factor two represents the interchange of the
two final--state pions). Some of the rescattering diagrams lead to an imaginary
part at threshold since they are evaluated at $\omega = 2M_\pi$ well above the
branch cut at $\omega_0 = M_\pi$

\item[$\bullet$] {\it Tree graphs with insertions from
${\cal L}_{\pi \pi}^{(4)}$}: These are the ones which are sensitive to the
pion--pion interaction at next--to--leading order which was studied in great
detail by Gasser and Leutwyler [5].

\item[$\bullet$] {\it Tree graphs with insertions from
${\cal L}_{\pi N}^{(2)}$}: These appear only with an extra suppression
factor of $1/m$ (otherwise they would have already shown up in the LETs,
Eq.(3)). The corresponding low--energy constants can be fixed by calculating
the subthreshold expansion of the elastic $\pi N$ scattering amplitudes and
comparing these to the empirical values as given e.g. by H\"ohler [6].

\item[$\bullet$] {\it Tree graphs with insertions from
${\cal L}_{\pi N}^{(3)}$}: These are in fact the ones which are the hardest to
pin down. Since there are not enough data to fix them all, they are estimated
in
Ref.[1] by resonance exchange (baryon excitations in the $s$--channel and
meson excitations in the $t$--channel). It turns out that for the threshold
amplitudes, only the excitation of the Roper $N^* (1440)$ is of relevance,
see also Ref.[7].

\end{itemize}

\noindent Before presenting results,
let me elaborate a bit on the Roper contribution.
While the $N^* N \pi$ vertex is fairly well known, it was shown in Ref.[1]
that for the $N^* N (\pi \pi)_S$ vertex there are effectively two {\it
different} lowest order couplings (written here non-relativistically),
\beq
{\cal L}_{N^* N \pi \pi} = c_1 \,M_\pi^2 \, N^* \,  {\vec{\pi}}^2 \, N
+ c_2 \, N^* \, (v_\mu \partial^\mu \vec{\pi} \,)^2 \, N
\eeq
where $v_\mu$ is the four--velocity of the heavy baryon. Both terms in Eq.(5)
have chiral dimension two. Since only the branching ratio $N^* \to N (\pi
\pi)_S$ is known, performing the phase space integration leads to a quadratic
form in the coupling constants $c_1$ and $c_2$, defining an ellipse.
 This ellipse is rather
elongated and thus leads to a large uncertainty in the sum $c_1+c_2$ since the
energy--dependence of the second coupling in Eq.(5) gives rise to a substantial
enhancement compared to the energy--independent coupling. It remains a
challenge to disentangle these coupling constants and further pin down
their numerical values. Furthermore, the $N^*$ contributes solely to $D_2$.

\vspace{0.5cm}

\noindent {\bf 3 $\quad$ RESULTS FOR THE $\pi N \to \pi \pi N$
                         THRESHOLD AMPLITUDES}

\vspace{0.3cm}

\noindent Consider first the amplitude $D_1$. Adding all theoretical
uncertainties in quadrature (for details, see Ref.[1]), one finds
\beq
D_1 = 2.65 \pm 0.24 \, \, {\rm fm}^3 \, \, \, ,
\eeq
which overlaps within one standard deviation with the empirical value, Eq.(4).
It is worth to note that the theoretical uncertainty is larger than the
empirical one. Most interesting, however, is the chiral expansion of $D_1$,
i.e. its separation into the contributions of order one, $M_\pi$ and $M_\pi^2$
(modulo logs),
\beq
D_1 = 1.59 \cdot ( \, 1.00 + 0.52 + 0.15 \, ) \, \, {\rm fm}^3 \, \, \, ,
\eeq
which shows a rapid convergence. The first three terms in the chiral
expansion behave like $1.59 \,\, {\rm exp}(0.52)$, so one expects the
correction at order $M_\pi^3$ to be very small. Note, however, that such an
estimate can not substitute for a full calculation to higher order.

In the case of $D_2$, matters are different. The central value including
all corrections to ${\cal O}(M_\pi^2)$ comes out surprisingly close to the
empirical value (cf. Eq.(4)),
\beq
D_2 = -9.06 \pm 1.05 \, \, {\rm fm}^3 \, \, \, ,
\eeq
where the large uncertainty is mostly due to the Roper as discussed before.
While one might content oneself with such a result, a closer look at the
chiral expansion
\beq
D_2 = -4.76 \cdot ( \, 1.00 + 0.42 + 0.48 \, ) \, \, {\rm fm}^3 \, \, \, ,
\eeq
reveals that there are still large corrections at ${\cal O}(M_\pi^2)$. So only
a calculation to (at least) one more order in the chiral expansion
(which is still one loop)
can clarify whether the nice agreement between theory and experiment for $D_2$
is accidental or not. Such a calculation is not yet available and can only
be performed if one clarifies before the nature and strengths of certain
resonance couplings entering at that order (like the Roper already discussed
but also the $N \Delta \pi \pi$ coupling and so on).

\vspace{0.5cm}

\noindent {\bf 4 $\quad$ RESULTS FOR THE S--WAVE $\pi \pi$ SCATTERING LENGTHS}

\vspace{0.3cm}

\noindent While the isospin two S--wave $\pi \pi$ scattering length is directly
proportional to $D_1$, the value for $a_0^0$ has to be extracted from the
combination $-2D_1 - 3D_2$. We can therefore expect to determine $a_0^2$
rather presicely whereas the resulting value for $a_0^0$ has to be considered
indicative due to the possible large higher order corrections in $D_2$. Adding
the theoretical and empirical uncertainties in quadrature, one finds
\beq
a_0^0 = 0.21 \pm 0.07 \, \, , \quad a_0^2 = -0.031 \pm 0.007 \, \, ,
\eeq
which both are in good agreement with the one--loop chiral perturbation theory
prediction of Gasser and Leutwyler [5],
\beq
a_0^{0, \, {\rm CHPT}} =  0.20 \pm 0.01 \, \, , \quad
a_0^{2, \, {\rm CHPT}} = -0.042 \pm 0.008 \, \, .
\eeq
Furthermore, the value for the combination $2a_0^0 - 5a_0^2 = 0.577 \pm 0.144$
is consistent with the so--called universal curve (UC),
$(2a_0^0 - 5a_0^2)_{\rm UC} = 0.614 \pm 0.028$. The numbers given here
supersede the previously determined $\pi \pi$ scattering lengths
from the threshold $\pi \pi N$ amplitudes based on
the Olsson--Turner model [8] which is only compatible with QCD for $\xi =0$
(since by now we know that the symmetry breaking is $\bar{3} \times 3$) and
thus amounts to the tree level approach.

\vspace{0.5cm}

\noindent {\bf 5 $\quad$ SHORT SUMMARY}

\vspace{0.3cm}

\noindent The reaction $\pi N \to \pi \pi N$ at threshold has been evaluated
within the framework of chiral perturbation theory to one loop accuracy [1].
It appears to be well suited for a precise determination
of the isospin two S--wave pion--pion scattering length $a_0^2$.
In case of the isospin zero scattering length $a_0^0$,
there are still large theoretical uncertainties related to interesting
baryon resonance physics which make an accurate determination very difficult.
However, it is worth noticing the stunning agreement of
the extracted value for $a_0^0$  with the one--loop CHPT prediction [5] --
is that merely an accident?  More theoretical effort is needed
to clarify this issue. At present, it appears that $K_{\ell 4}$--decays or
pionic molecules are better candidates to precisely pin down $a_0^0$.

\vspace{0.5cm}

\noindent {\bf 6 $\quad$ ACKNOWLEDGEMENTS}

\vspace{0.3cm}

\noindent The results presented here have been obtained in a collaboration
with V\'eronique Bernard and Norbert Kaiser to whom I express my deep
gratitude. Many thanks also to the 'generalizers' from Orsay for some
very valuable comments.

\vspace{0.8cm}
\noindent{\bf References}

\bigskip

\noindent 1. V. Bernard, N. Kaiser, and Ulf-G. Mei{\ss}ner, {\it Nucl. Phys.\/}
{\bf B} (1995) in print.

\smallskip

\noindent 2. V. Bernard, N. Kaiser and Ulf-G. Mei{\ss}ner,
{\it Phys. Lett.\/} {\bf B332}, 415 (1994);

{\it ibid} (E) {\bf B338}, 520 (1994).

\smallskip
\noindent 3. V. Bernard, N. Kaiser, and Ulf-G. Mei{\ss}ner,
{\it Int. J. Mod. Phys.\/} {\bf E4}, 193 (1995).

\smallskip

\noindent 4. H. Burkhard  and J. Lowe,  {\it Phys. Rev. Lett.}
{\bf 67}, 2622 (1991).

\smallskip

\noindent 5. J. Gasser and H. Leutwyler,  {\it Ann. Phys.} (NY)
{\bf 158}, 142 (1984).

\smallskip

\noindent 6. G. H\"ohler, in Landolt-B\"ornstein, vol.9b2, H. Schopper (ed.),
Springer, Berlin (1983).

\smallskip

\noindent 7. D.M. Manley et al., {\it Phys. Rev.} {\bf D30}, 904 (1984).

\smallskip

\noindent 8. M.G. Olsson and Leaf Turner, {\it Phys. Rev. Lett.}
{\bf 20}, 1127 (1968).

\smallskip

\end{document}